\documentclass[11pt]{llncs}
\usepackage{times}
\usepackage{amsfonts}
\usepackage{amsmath}
\usepackage{amssymb}
\usepackage[pdftex]{graphicx}

\usepackage[small,it]{caption} 

\newcommand{\ms}[1]{\ensuremath{\mathsf{#1}}}

\newcommand{\argmin}{\operatornamewithlimits{arg\ min}}

\def\reals{{\mathbb R}}

\newcommand{\ver}{{\ms{V}}}
\newcommand{\edge}{{\ms{E}}}
\renewcommand{\deg}{{\ms{d}}}

\newcommand{\uni}{U}

\newcommand{\energy}{{\mathcal{E}}}

\newcommand{\ham}{{\mathcal{H}}}

\newcommand{\emb}{{\ms{emb}}}

\newcommand{\Gemb}{{G_{\ms{emb}}}}

\newcommand{\AH}{{\sc Triad}}

\def\final{0} 
\ifnum\final=1  
\newcommand{\vnote}[1]{[{\small Vicky: \bf #1}]\marginpar{*}}

\newcommand{\sidecomment}[1]{\marginpar{\tiny #1}}
\else 
\newcommand{\vnote}[1]{}
\newcommand{\sidecomment}[1]{}
\fi  

\begin{document}
\title{ Minor-embedding in adiabatic quantum computation:
II. Minor-universal graph design}
\author{Vicky Choi\inst{}}
\institute{Department of Computer Science,
Virginia Tech}

\maketitle       
\begin{abstract}
In \cite{minor-embedding},  we introduced the notion of
minor-embedding in adiabatic quantum optimization.
A minor-embedding of a graph $G$ in a quantum hardware graph $\uni$ is
a subgraph of $U$ such that $G$ can be obtained from it by contracting edges. 
In this paper, we describe  the intertwined adiabatic quantum architecture design problem, which is to construct 
a hardware graph $U$ that satisfies all known physical constraints and, at the same time, permits an efficient 
minor-embedding algorithm.  We illustrate an optimal complete-graph-minor hardware graph. 
Given a family $\mathcal{F}$ of graphs, a (host) graph $\uni$ is called $\mathcal{F}$-minor-universal if
for each graph $G$ in $\mathcal{F}$, $\uni$ contains a minor-embedding
of $G$.  The problem for 
designing a
${\mathcal{F}}$-minor-universal hardware graph $\uni_{\ms{sparse}}$ in which
${\mathcal{F}}$
consists of a family of sparse graphs (e.g., bounded degree graphs) is open.
\end{abstract}

\section{Introduction}
We  introduced the notion of minor-embedding in adiabatic quantum
optimization in \cite{minor-embedding}.
In particular, we showed that the NP-hard quadratic unconstrained binary optimization problem on a graph $G$
can be solved in an adiabatic quantum computer that implements the spin-1/2 Ising Hamiltonian, by reduction
through minor-embedding of $G$ in the quantum hardware graph $U$.
We proved the correctness of the minor-embedding reduction and solved
a related
parameter setting problem in \cite{minor-embedding}.
In this paper, we discuss the intertwined adiabatic quantum architecture design problem, which is to construct 
a hardware graph $U$ that satisfies all known physical constraints and, at the same time, permits an efficient 
minor-embedding algorithm.

This paper is organized as follows. 
In Section~\ref{sec:AQC}, we review adiabatic quantum computation, and 
describe an
adiabatic quantum architecture that is implemented with
 superconducting devices and the imposing physical constraints.
In Section~\ref{sec:minor-embed},
we recall the minor-embedding in AQC.
In Section~\ref{sec:triad}, we describe an optimal complete-graph
minor hardware graph. 
In Section~\ref{sec:sparse}, we discuss the open problem for designing sparse-graph minor universal hardware graphs  
and the 
related work in literature.

\section{Adiabatic Quantum Optimization and Adiabatic Quantum
  Hardware Graph}
\label{sec:AQC}
Adiabatic quantum computation (AQC)
was proposed by Farhi~et~al.~\cite{FGGS00,FGGLLP01} in 2000 
as an alternative quantum paradigm to solve NP-hard 
optimization problems, which are believed to be classically intractable.
Later, it was shown by Aharonov~et~al.~\cite{ADKLLR04} that AQC is not
just limited to optimization problems, and is polynomially equivalent to 
 conventional quantum computation (quantum circuit model). 
 In this
paper, we will focus on  {\em quantum adiabatic optimization}, 
in which the final Hamiltonian is a 
diagonal matrix in the computational basis. 
In particular, we restrict to a subclass of Hamiltonians, known as {\em Ising Hamiltonians}:
\begin{equation}
\ham_{\ms{Ising}} = \sum_{i \in \ver(G)} h_i \sigma^z_i + \sum_{ij \in \edge(G)} J_{ij}
\sigma^z_i \sigma^z_j.
\label{eqn:Ising-ham}
\end{equation}
where $\ver(G)$ ($\edge(G)$ resp.) is the vertex set (edge set resp.) of $G$, $h_i, J_{ij} \in \reals$, 
and $\sigma_i^z = I \otimes I \otimes \ldots \otimes \sigma^z \otimes
\ldots \otimes I$ (the Pauli matrix $\sigma^z$ is in the $i$th position), similarly for
$\sigma^z_i\sigma^z_j$.




The eigenvalues and the corresponding eigenstates of
$\ham_{\ms{Ising}}$ are
encoded in the following energy function:
\begin{eqnarray}
\energy(s_1,\ldots, s_n) = \sum_{i \in \ver(G)} h_i s_i + \sum_{ij \in\edge(G)} J_{ij}
      s_is_j
\label{eqn:OE}
\end{eqnarray}
where $s_i \in \{-1,+1\}$, called a {\em spin}.
In particular, the smallest eigenvalue of $\ham_{\ms{Ising}}$ corresponding to the minimum
 of $\energy$, and $\argmin \energy$ corresponds to its eigenvector (called {\em ground
  state}) of $\ham_{\ms{Ising}}$. 
Hereafter, we refer the problem of
finding the minimum 
energy of the Ising model or
equivalently the ground state of Ising Hamiltonian as the {\em Ising problem}.

\subsection{Adiabatic Quantum Algorithm}
An adiabatic quantum algorithm is described by a system Hamiltonian
\begin{equation*}
\ham(t) = (1-s(t))\ham_{\ms{init}} + s(t) \ham_{\ms{final}}
\end{equation*}
for $t \in [0,T]$, $s(0)=0$, $s(T)=1$.
There are three ingredients of $\ham(.)$: 
(1) initial Hamiltonian: $\ham(0)=\ham_{\ms{init}}$;
(2) final Hamiltonian:  $\ham(T)=\ham_{\ms{final}}$;
and (3) evolution path: $s : [0,T] \longrightarrow [0,1]$, e.g., $s(t)=\frac{t}{T}$.
$\ham(.)$ is an adiabatic algorithm for an optimization problem if we encode the problem into the final 
Hamiltonian $\ham_{\ms{final}}$ such that the ground state of $\ham_{\ms{final}}$ corresponds to the answer to
the problem. The initial Hamiltonian $\ham_{\ms{init}}$ is chosen to be non-commutative with $\ham_{\ms{final}}$
and its ground state must be known  and experimentally constructable, e.g.,  $\ham_{\ms{inital}} = -\sum_{i \in \ver(G)} \Delta_i \sigma_i^x$.
Here $T$ is the running time of the algorithm.
According to the adiabatic theorem, if $\ham(t)$ evolves ``slowly'' enough, or equivalently, if $T$ is large enough,
which is determined by the minimum spectral gap (the difference between the two lowest energy levels) of
the system Hamiltonian, 
 the system remains at the ground state of $\ham(t)$, and consequently, ground state of $\ham(T)=\ham_{\ms{final}}$ gives the solution to the problem. 

Therefore, if we set $\ham_{\ms{final}}$ to be $\ham_{\ms{Ising}}$, $\ham(.)$ is an adiabatic 
algorithm for the Ising problem.  In physics, 
\begin{equation*}
\ham(t) = (1-s(t)) \left(\sum_{i \in \ver(G)} \Delta_i \sigma_i^x\right)  + s(t) \left( \sum_{i \in \ver(G)} h_i \sigma^z_i + \sum_{ij \in \edge(G)} J_{ij}
\sigma^z_i \sigma^z_j \right)
\end{equation*}
is known as the {\em Ising model in a transverse field}.

\subsection{Superconducting Architecture for AQC}
\begin{figure}[h]
$$
\begin{array}{cc}
\includegraphics[width=0.4\textwidth]{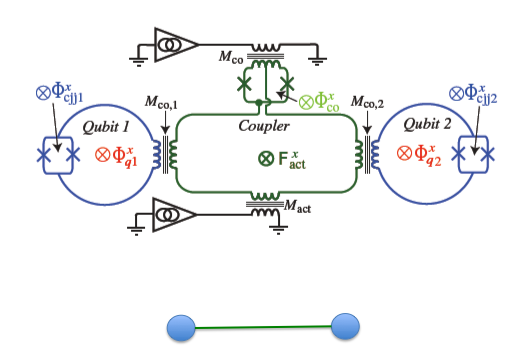} & 
\includegraphics[width=0.3\textwidth,angle=90]{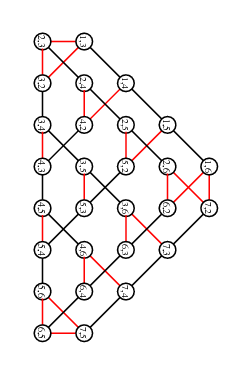}\\
(a) & (b)
\end{array}
$$
\caption{(a) Schematic of superconducting based implementation of qubit-coupler-qubit~\cite{harris-2009}.
(b) An example hardware graph of $28$ qubits~\cite{d3-triad}. Each qubit
  is coupled with exactly 3 other qubits.}
\label{fig:physical-qubit}
\label{fig:d3-triad}
\end{figure}

A scalable superconducting architecture for adiabatic quantum
  computation that implements the Hamiltonian of Ising model in a transverse field was initially proposed by Kaminsky~et~al.~\cite{KLO04}. 
D-Wave Systems Inc.~\cite{Dwave} is building such a superconducting quantum processor.
In particular, the quantum 
architecture is based on superconducting flux qubits connected via tunable coupling devices. 
See Figure~\ref{fig:physical-qubit}$(a)$ for the schematic of physical qubit-coupler-qubit, and \cite{harris-2009} for its design and experimental results.

Topologically, an adiabatic quantum hardware architecture can be viewed as an undirected
graph $\uni$ with weighted vertices and weighted edges. See
Figure~\ref{fig:d3-triad}$(b)$ for an example.
Let $\uni$ denote the quantum hardware graph.
Each vertex $i \in \ver(\uni)$ corresponds to a qubit, and each edge $ij
\in \edge(\uni)$ corresponds to a coupler between qubit $i$ and qubit
$j$. 
In the following, we will use qubit and vertex, and coupler and edge
interchangeably when there is no confusion.
There are two weights, $h_i$ (called the qubit {\em bias}) and $\Delta_i$
(called the {\em tunneling amplitude}), associated
with each qubit $i$. There is a weight $J_{ij}$ (called the coupler {\em
  strength})
 associated with each coupler
$ij$. In general, these weights are functions of time, i.e., they vary
over  time, e.g., $h_i(t)$.

\subsection{Physical Constraints}
There are some known physical constraints on the quantum hardware graph (superconductor based
design).  
In particular, there is a {\em degree-constraint} in that
each qubit can have at most a constant number 
of couplers. The coupler (or edge) length can not be ``too long'' (that is,  all neighbor qubits are within
a bounded distance). Note that the wire of a qubit can be ``stretched''. The shape of each qubit does not need to be a small circle.
In other words, an adiabatic quantum hardware graph is a
bounded-degree, edge-length bounded {\em geometric graph} (or known as {\em layout}). Notice that crossing is allowed
(i.e., it can be a non-planar graph).

\section{Minor-Embedding in AQC}
\label{sec:minor-embed}

\begin{figure}[h]
\centering{
\includegraphics[width=0.45\textwidth]{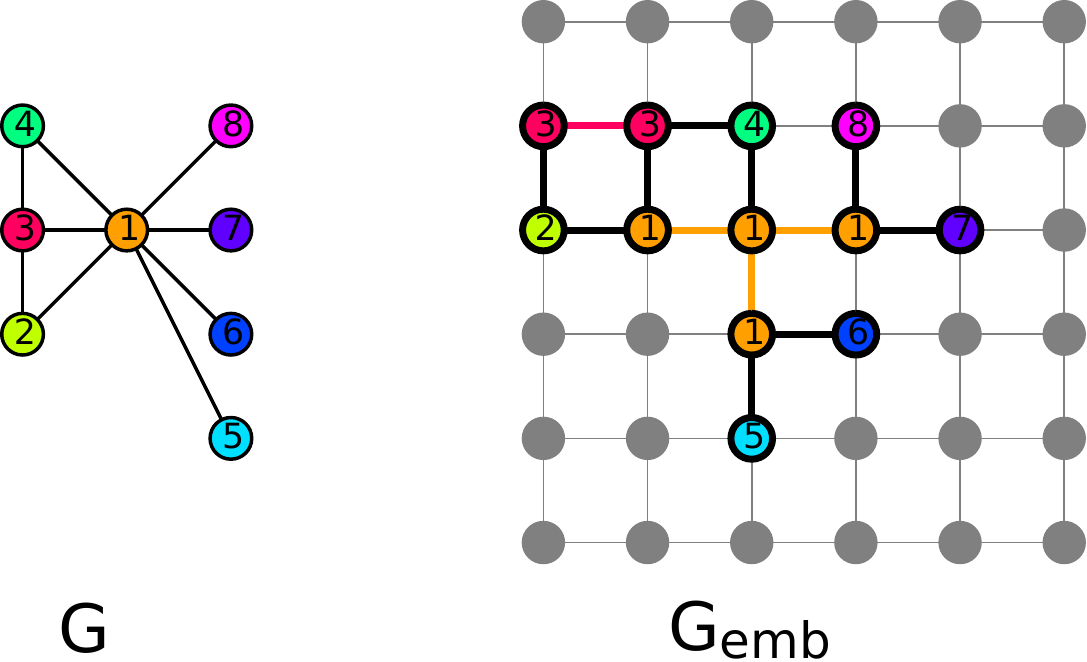}
} 
\caption{$\Gemb$(right) is a minor-embedding of $G$(left) in the square
  lattice $\uni$. Each vertex (called a logical qubit) of $G$ is mapped to a
  (connected) subtree of (same color/label) vertices (called physical qubits) of $\uni$. 
$G$ is called a (graph) minor of $\uni$.}
\label{fig:example-embedding}
\end{figure}

Given an adiabatic quantum computer that implements the Ising Hamiltonian, 
one can thus solve an Ising problem on a graph $G$, 
if $G$ can be embedded as a subgraph of the quantum hardware graph
$\uni$. 
As mentioned above,  there are physical constraints on the hardware
graph $\uni$. In particular, the {\em degree-constraint} of a qubit implies that 
the graphs that can be
solved on a given hardware graph $\uni$ through
subgraph embedding must also be degree-bounded. 
So what to do if $G$ is not a subgraph of the hardware graph $\uni$?
Kaminsky~et~al.~\cite{KL02,KLO04}
observed and proposed that one can embed $G$ in $U$ through
ferromagnetic coupling  to solve 
Maximum Independent Set (MIS) problem (which is a special case of the Ising problem)
of planar cubic graphs (regular graphs of
degree-3)
on an adiabatic quantum computer. 
In particular, they proposed an $n \times n$
square lattice as a scalable hardware architecture on which all $n/3$-vertex planar cubic graphs are
embeddable.  
The notion of embedding here follows naturally from physicists' intuition
that 
each {\em logical qubit} (corresponding to a vertex in the input
graph) is mapped to a subtree of {\em physical qubits} (corresponding to vertices in
the hardware graph) that are ferromagnetically coupled such 
that each
subtree of physical qubits acts like a single logical qubit.
For example, in Figure~\ref{fig:example-embedding}, the logical qubit $1$
(in orange color) of the graph $G$ is mapped to a subtree of physical
qubits (labeled $1$) of the square lattice. 
Informally, a minor-embedding $\Gemb$ of a graph $G$ in the hardware graph
$\uni$ is a subgraph of $\uni$ such that $\Gemb$ is an ``expansion'' of $G$
by replacing each vertex of $G$ with a (connected) subtree of $\uni$, or
equivalently, $G$ can be obtained from $\Gemb$ by contracting edges (same color
in Figure~\ref{fig:example-embedding}). In
graph theory, $G$ is called a (graph) {\em minor} of $\uni$. (see for
example~\cite{Distel05}). 

We now formally define minor-embedding.
\begin{definition}
Let $\uni$ be a fixed hardware graph. Given $G$, 
the {\em minor-embedding} of $G$ is
defined by 
$$\phi: G \longrightarrow \uni$$ such that
\begin{itemize}
\item each vertex in $\ver(G)$ is mapped to a connected subtree $T_i$ of $\uni$;
\item there exists a map $\tau:
  \ver(G) \times \ver(G) \longrightarrow \edge(\uni)$
such that 
for each $ij \in \edge(G)$, there are corresponding $i_{\tau(i,j)}
  \in \ver(T_i)$ and $j_{\tau(j,i)} \in \ver(T_j)$
  with  $i_{\tau(i,j)}j_{\tau(j,i)} \in \edge(\uni)$.
\end{itemize}  
Given $G$, if $\phi$ exists, we say that $G$ is {\em
  embeddable}
 in $\uni$. 
When $\phi$ is clear from the context, we denote the minor-embedding
  $\phi(G)$ of $G$
  by $\Gemb$\footnote{With slight abuse of terminology, $\Gemb$ is also referred as $G$-minor.}.  
\end{definition} 

See Figure~\ref{fig:example-embedding} for an example.

In particular, there are two special cases of minor-embedding:
\begin{itemize}
\item{\em Subgraph-embedding:} Each $T_i$ consists of a single vertex in
  $\uni$. That is, $G$ is isomorphic to $\Gemb$ (a subgraph of $\uni$).
\item{\em Topological-minor-embedding:} Each $T_i$ is a chain (or path) of vertices in $\uni$.
\end{itemize}
Remark: The embedding in \cite{KL02,KLO04} is the {\em topological-minor} embedding.

In~\cite{minor-embedding},  we have shown that the NP-hard quadratic unconstrained binary optimization problem~\cite{BH02,BHT06} (which is equivalent to the Ising problem) on a graph $G$
can be solved in an adiabatic quantum computer that implements the spin-1/2 Ising Hamiltonian, by reduction
through minor-embedding of $G$ in the quantum hardware graph $U$.
By {\em reduction
through minor-embedding}, we mean that one can reduce the original Ising Hamiltonian on the input
graph $G$ to the embedded Ising Hamiltonian
$\ham^{\emb}$ on its minor-embedding $\Gemb$, i.e., the solution to the
embedded Ising Hamiltonian gives rise to the solution to the original
Ising Hamiltonian.
We proved the correctness of the minor-embedding reduction.
There are two components to
the reduction: {\em embedding} and {\em parameter setting}.
The embedding problem is to find a minor-embedding
  $\Gemb$ 
  of a graph $G$ in $\uni$. 
The parameter setting problem is to set the
  corresponding parameters, qubit bias and coupler strengths, of the embedded
  Ising Hamiltonian.
In \cite{minor-embedding}, we solved the parameter setting problem. 
The embedding problem, though,  is dependent on the hardware graph design problem discussed in the following sections.

\section{\AH{}: Optimal Hardware Graph for Embedding Complete
  Graph $K_n$}
\label{sec:triad}
\begin{figure}
$$
  \begin{array}{cc}
    \includegraphics[width=0.3\textwidth]{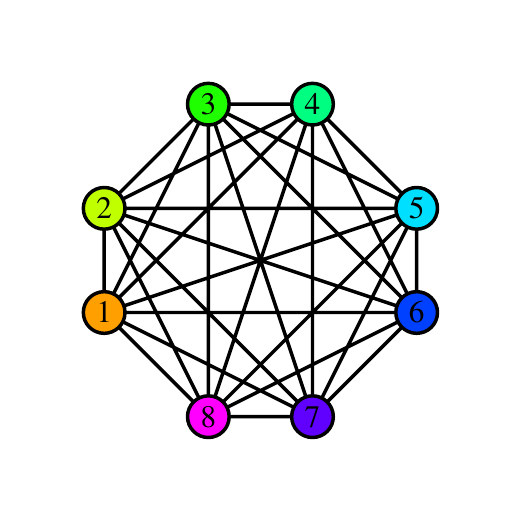}
& \includegraphics[width=0.3\textwidth,angle=90]{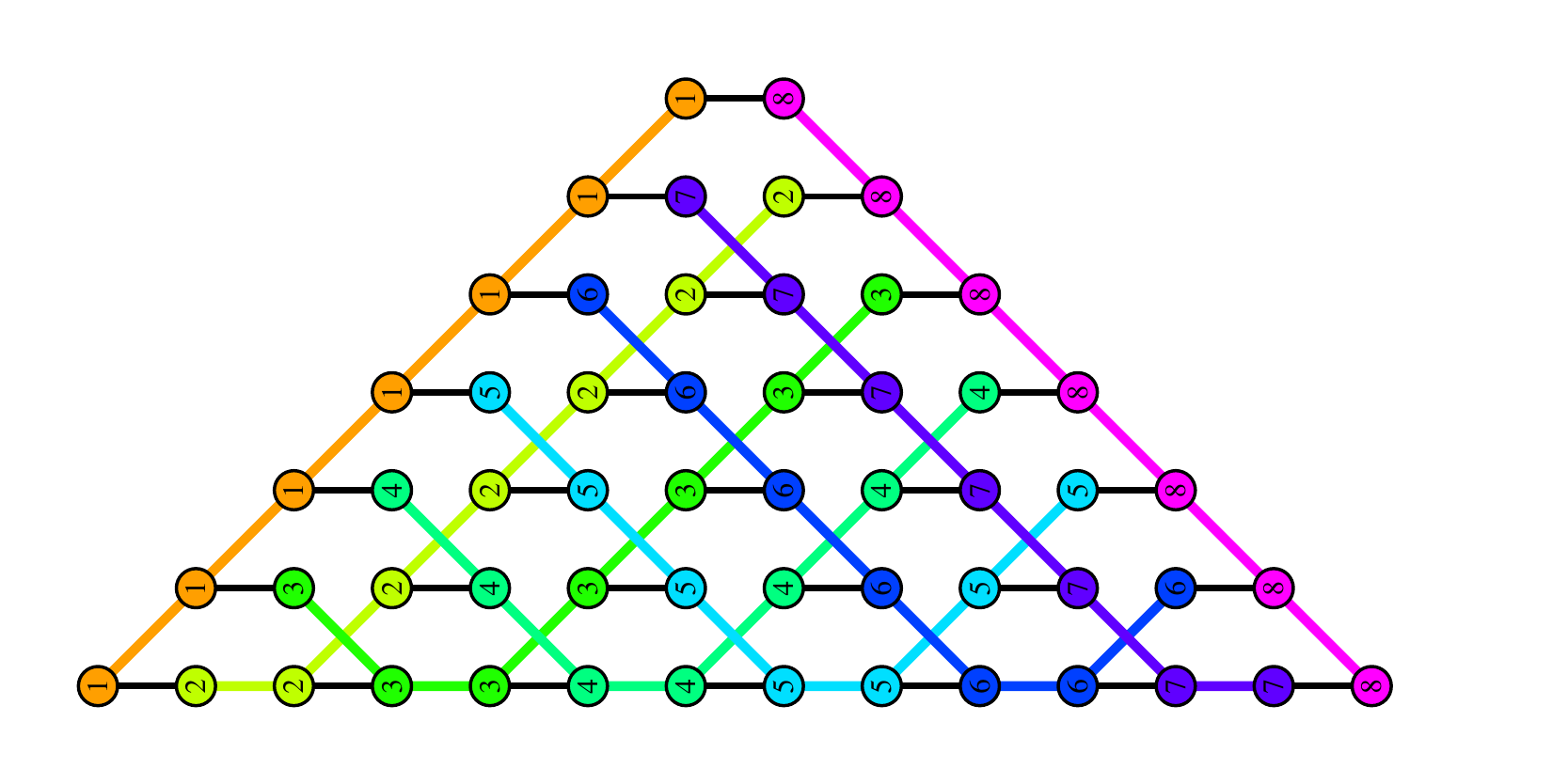}
  \end{array}
$$
\caption{Left, $K_8$. Right, a triangular layout of a $K_8$-minor. Each vertex of $K_8$ is mapped to a chain of $7$ ``virtual'' vertices (with the same color).}
\label{fig:triad}
\end{figure}

In this section, we describe 
a $K_n$-minor hardware graph, where
$K_n$ is a complete graph of $n$ vertices. 
A triangular layout of a $K_n$-minor graph~\cite{triad-patent}, called {\AH{}}, is shown in Figure~\ref{fig:triad}. 

\subsection{Construction of \AH{}}
\begin{figure}[h]
$$
\begin{array}{ccc}
K_3 & K_4 &K_5\\
\includegraphics[width=0.15\textwidth]{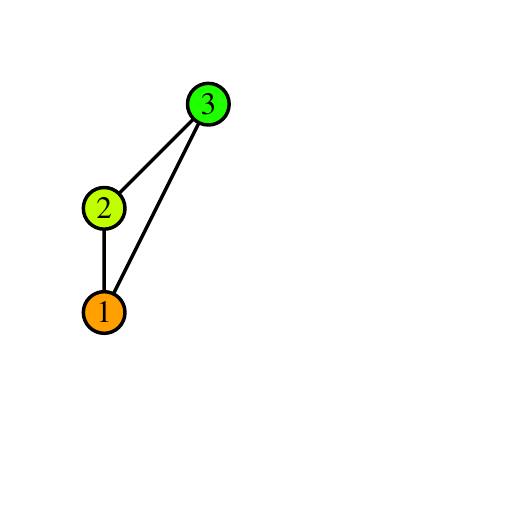} &
\includegraphics[width=0.15\textwidth]{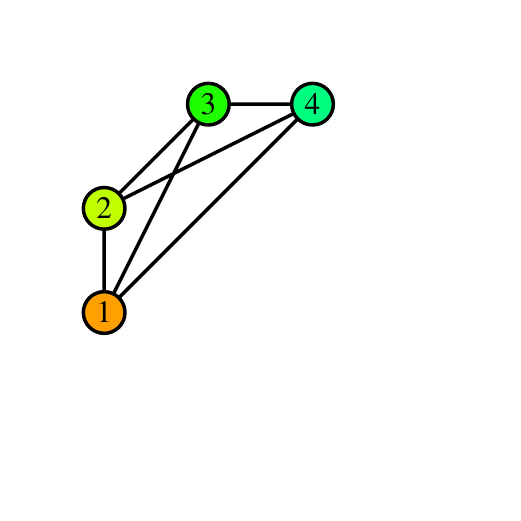} &
\includegraphics[width=0.15\textwidth]{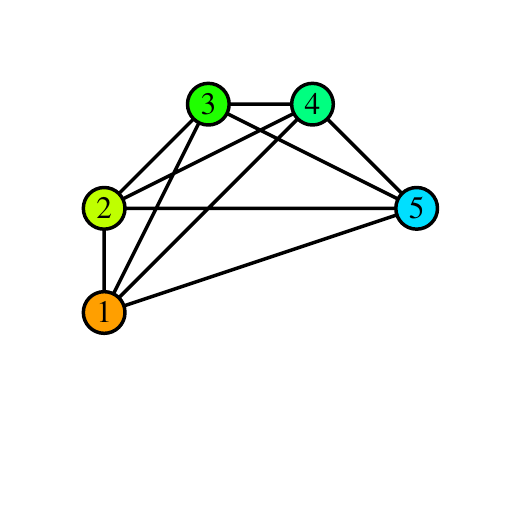} \\
\includegraphics[width=0.15\textwidth,angle=90]{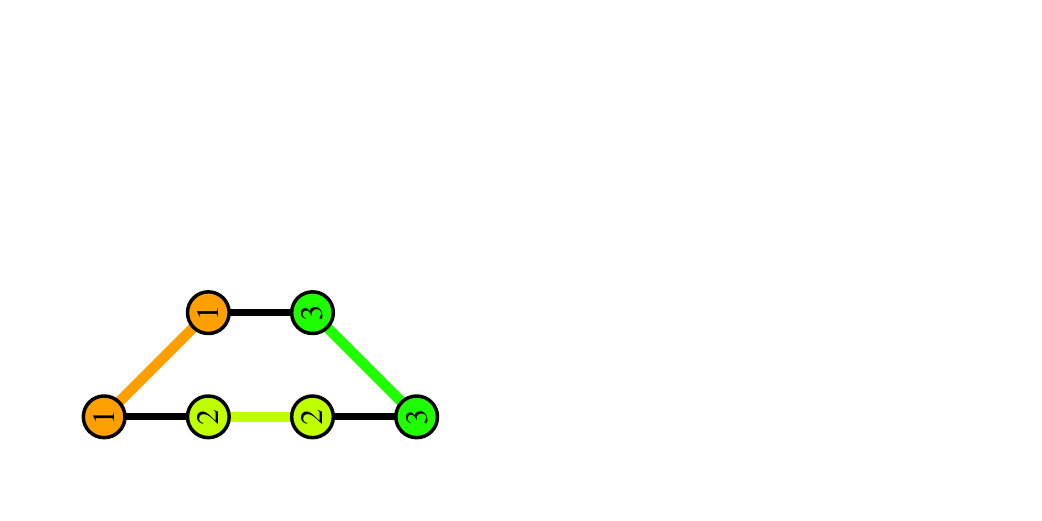} &
\includegraphics[width=0.15\textwidth,angle=90]{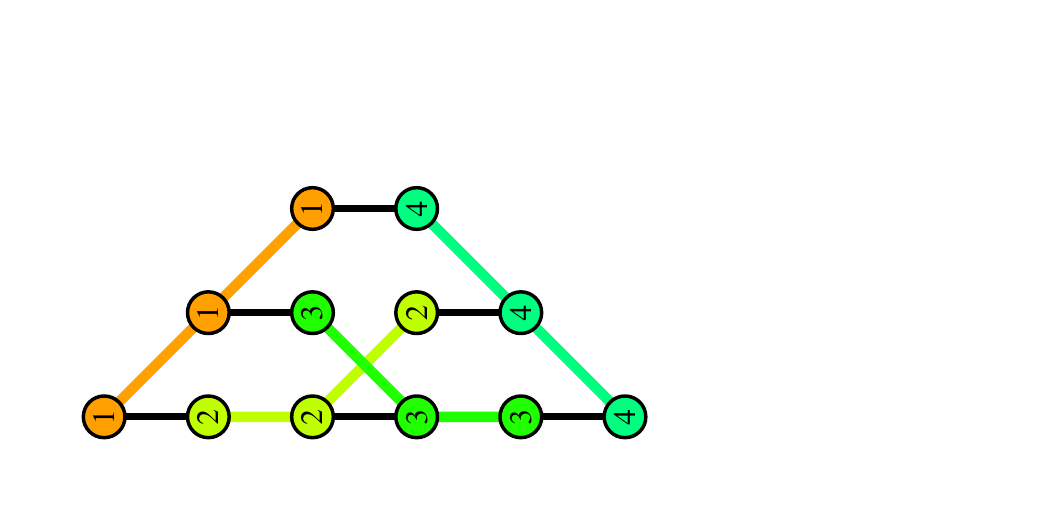} &
\includegraphics[width=0.15\textwidth,angle=90]{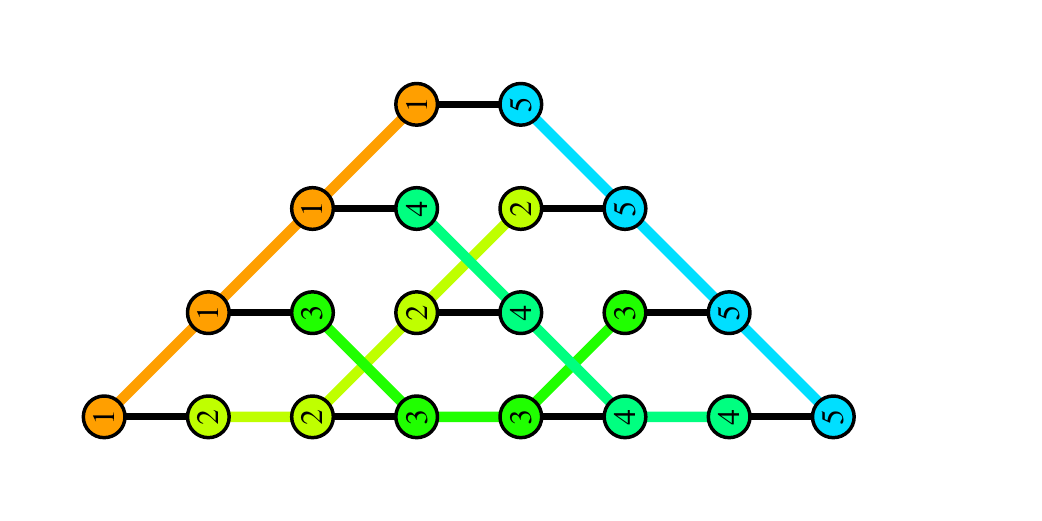}
\end{array}
$$
\caption{Incremental construction: $K_3 \Longrightarrow K_4
  \Longrightarrow K_5$. 
Starting with the basis layout of $K_3$ where each vertex is represented by a chain of two (which is the degree)  virtual  vertices.
To construct the layout of $K_4$ from $K_3$: extend each chain in the north-east direction by one new virtual vertex; place a chain of three virtual vertices for vertex $4$ accordingly to achieve the needed adjacency.
Inductively, we construct $K_n$ from $K_{n-1}$. 
}
\label{fig:construction}
\end{figure}

The idea behind the construction of \AH{} is to map each vertex  of $K_n$
to a chain of $n-1$ ``virtual'' vertices. The inductive construction is illustrated in Figure~\ref{fig:construction}.

\subsection{Decomposition of \AH{}}
\begin{figure}[h]
$$
\begin{array}{cc}
  \includegraphics[width=0.55\textwidth]{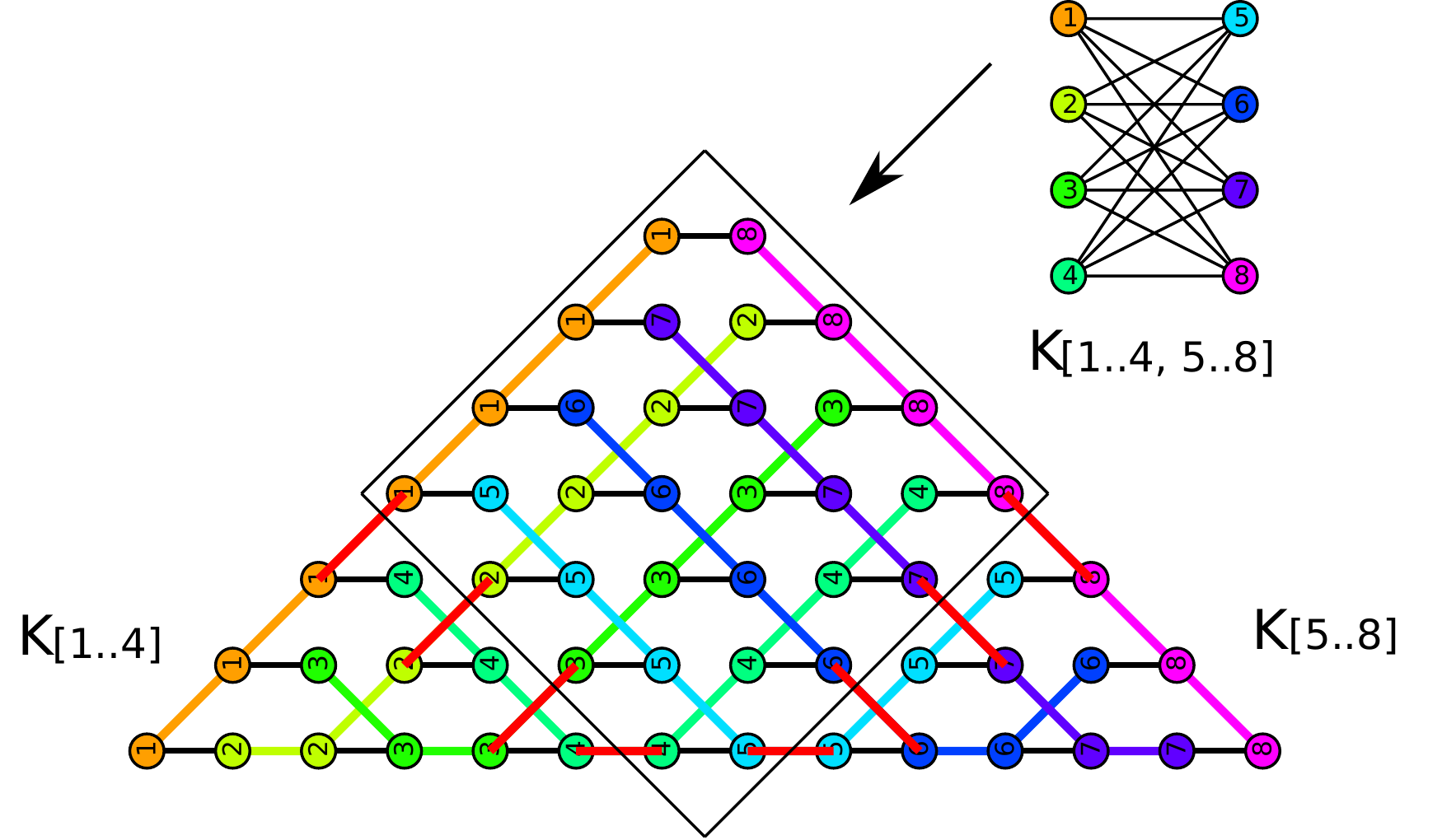} &
\includegraphics[width=0.45\textwidth]{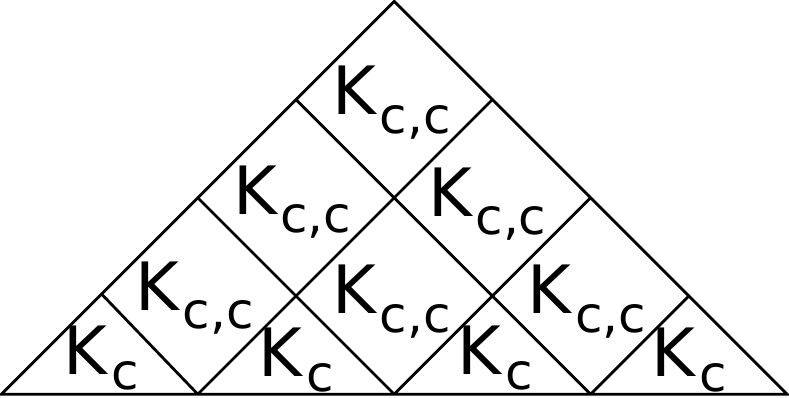}\\
(a) & (b)
\end{array}
$$
\caption{(a) Chop (indicated by the red line) each chain of $7$ virtual vertices into two physical qubits: one consists of first 3 virtual vertices and the other consists of remaining
4 virtual vertices. The result is a decomposition: $K_8= K_{[1..4]}  \cup
 K_{[1..4],[5..8]} \cup K_{[5..8]} $.
(b) A complete graph $K_n$ (assume $n=2^{ck}$) can be represented as a partition of
  complete graph $K_c$'s and complete bipartite graph $K_{c,c}$'s. }
\label{fig:decomp}
\label{fig:decomp2}
\end{figure}

Suppose the available degree (the number of allowed couplers) of a physical qubit is $\deg$. 
The idea is to chop the $n-1$ virtual vertices into
$\lceil\frac{n-3}{\deg-2}\rceil$ physical qubits\footnote{because each non-terminal physical qubit requires at least  two couplers to connect to its adjacent physical qubits; there are at least two terminal physical qubits which require one coupler.}.  For example, suppose $n=8$ and $\deg=6$,  then it will require two physical qubits: we ``chop'' the chain 
of $7$ virtual vertices into two physical qubits: one consists of 3 virtual vertices and the other consists of 
4 virtual vertices.  See Figure~\ref{fig:decomp}$(a)$ for the illustration.
The result of such a chopping is a decomposition of $K_n$. 
Alternatively, we can view a complete graph $K_n$ as a
combination (disjoint union) of two complete subgraphs of $K_{\frac{n}{2}}$ and a complete
bipartite graph of $K_{\frac{n}{2},\frac{n}{2}}$.
Inductively, a complete graph $K_n$ (assume $n=2^{ck}$) can be represented as a partition of
  complete graph $K_c$'s and complete bipartite graph $K_{c,c}$'s, as illustrated in Figure~\ref{fig:decomp2}$(b)$. 
Notice that $K_{c}$ is a minor of $K_{c,c}$. For example, if $\deg=6$,
one can thus use $K_{4,4}$ as
  the basic unit for building \AH{}.
By construction, it is easy to see that \AH{}  is optimized for embedding a complete graph $K_n$.
Each logical qubit (vertex) of $K_n$ requires at least $\lceil \frac{n-3}{\deg-2} \rceil$ physical qubits. 
That is,  the hardware graph needs to have $\Omega(n^2/\deg)$ physical qubits.

In short, \AH{} satisfies all known physical constraints and  admits a simple and
efficient embedding (as illustrated in Figure~\ref{fig:triad}) of 
 $K_n$, which thus allows for embedding any graph
of $n$ vertices efficiently. 
Since qubits
and couplers are scarce, we are interested in designing a sparse-graph-minor hardware graph.

\section{Adiabatic Quantum Architecture $\uni_{\ms{sparse}}$  Design Problem}
\label{sec:sparse}
First, we introduce the terminology of minor-universal graph.
\begin{definition}
      Given a family $\mathcal{F}$ of graphs, a (host) graph $\uni$ is called
      {\em $\mathcal{F}$-minor-universal} if for any graph $G \in \mathcal{F}$, 
there exists a minor-embedding of $G$ in $\uni$. 
\end{definition}
Let ${\mathcal{F}}$
consist of a set of sparse graphs (e.g., bounded degree graphs).
It is also desirable that these sparse graphs are the underlying graphs of some classically hard instances. 
We are interested in 
designing a ${\mathcal{F}}$-minor-universal graph {\em $\uni_{\ms{sparse}}$} that is as small as possible 
(in terms of number of  qubits and number of couplers ) subject
      to    both physical constraints and embedding constraint. That is, it satisfies all known physical constraints and, at the same time, a minor-embedding can be efficiently computed.

\subsection{Related Work and Discussion}
The name ``minor-universal'' was naturally adopted from the
``universal graph''. There are many studies on the construction of an universal graph
(which is a special case of a minor-universal graph),
see \cite{alon-universal} and references therein. 
Many of these work are based on the expanders which seem to have the desired properties --- sparse 
and yet highly connected. However, to the best of our knowledge, all the current known explicitly constructable expanders
are {\em topological} graphs and require long edge lengths, while what we need  is a {\em geometric} expander.

Minors are well-studied in graph theory, see for
example~\cite{Distel05}. 
Given a {\em fixed} graph $G$, there are algorithms that find a minor-embedding of $G$ in
$\uni$ in polynomial time of size of $\uni$, from the pioneering
$O(|\ver(\uni)|^3)$ time algorithm by Robertson and Seymour~\cite{RS95} to the recent nearly linear
time algorithm of B.~Reed. However, it is worthwhile to reiterate that these
algorithms are for {\em fixed} $G$, and their running times are {\em exponential} in the
size of $G$. 
Here the minor-embedding problem is to find a minor-embedding of $G$ (for any
given $G$) while
fixing $\uni$. 
To the best of our knowledge, the
only known work related to our minor-embedding problem was by Kleinberg and Rubinfeld~\cite{KR96}, in
which they showed that there is a randomized polynomial algorithm, based on a
random walk, to find a minor-embedding in a given degree-bounded
expander. However, as discussed above, the known expanders  do not
satisfy the physical constraints required.

Our embedding problem might appear similar to the embedding
  problem from parallel architecture studies. However, besides the
  different physical constraints for the design of architectures, the requirements are very
  different. In particular, in our embedding problem, we do not allow for
  {\em load} $>1$, which is the maximum number of logical qubits mapped to
  a single physical qubit. 
  Also, we require {\em dilation}, which is the
  maximum number of stretched edges (through other qubits), to be exactly 1.
However, all of the existing research on embedding problems for parallel
  processors~\cite{Leighton-book},
at least one of the conditions  is violated (namely either load
  $>1$ or dilation $>1$).

We remark that the {\em treewidth}  of $\uni_{\ms{sparse}}$ needs to necessarily large ($\omega(\log n)$) for
otherwise the dynamic programming over  the tree-decomposition of $\uni_{\ms{sparse}}$
 would be able to
solve the problem in $O(exp({\ms{tw}(\uni_{\ms{sparse}})}))$ time (while polynomial in the
size of the input size)~\cite{tree-decomposition},
once the
embedding is given.  For the quantum circuit model,  it was shown by Markov \& Shi~\cite{markov-2008} that
a quantum circuit with $n$ gates whose underlying graph has treewidth $\ms{tw}$ can be simulated in  
$O(poly(n) exp(\ms{tw}))$ time.

How does a minor-embedding affect the efficiency of an adiabatic algorithm? and how do we measure the goodness of an embedding? 
These problems in turn relate to the running time or
  complexity of quantum
  adiabatic algorithms.
  Recall that according to the adiabatic theorem, the running time of an
  adiabatic algorithm depends on the
minimum spectral gap of the system Hamiltonian, which however is in general difficult to
compute analytically.
In order to address the time complexity of an adiabatic algorithm, one will also need to specify the initial
Hamiltonian. 
The effect of the embedding  and its consequential initial
Hamiltonian
on the complexity of an adiabatic algorithm remains to be
investigated.

\subsection*{Acknowledgment}
Most of this work was done while the author was working at D-Wave Systems Inc. I would like 
to thank my colleagues from whom I learn to derive this work.
Thanks also go to David Kirkpatrick for the encouragement, advice and discussion, and Bill Kaminsky for the comments.


{\tiny

}
\end{document}